\documentclass[twoside, epsfig]{article}

\input oejv2.sty

\begin{document}

\OEJVhead{01 2024}
\OEJVtitle{Long term optical variability of two HMXB}
\OEJVauth{Roberto Nesci,$^1$; Mariateresa Fiocchi,$^1$; Antonio Vagnozzi,$^2$}
\OEJVinst{INAF/IAPS, via Fosso del Cavaliere 100, 00133, Roma, Italy, {\tt \href{mailto:email}{roberto.nesci@inaf.it}}}
\OEJVinst{MPC589, Via S.Lucia n.68, I-05039, Stroncone (TR), Italy}

\OEJVabstract{We discuss the optical light curves of two Be X-ray Binaries, IGR J06074+2205 and SAX J2103.5+4545 recovered from the ATLAS, ZTF and ASAS-SN databases. Both sources show long term optical variability of 620 and 420 days respectively, with color redder when brighter. We suggest that this is due to the precession of the circumstellar disk.  Another possibility is the propagation of a density wave in the disk. We remark that only these two sources show a large amplitude, periodic optical variability out of a sample of 16 well studied High Mass X-ray Binaries (HMXB) with a Main Sequence primary star.
}

\begintext

\section{Introduction}\label{secintro} 
IGR J0607+2205 is a B0.5Ve HMXB discovered by INTEGRAL in 2003 \citep{ATel223} which recently underwent several X-ray outbursts 
\citep{ATel15294, ATel16267, ATel16277, ATel16278, ATel16307}, after many years of quiescence. From these outbursts a likely orbital periodicity of 80 days was inferred \citep{ATel16351, ATel16394}.

We searched therefore this source in some public databases of optical all sky monitoring, namely ASAS-SN \citep{Sha14}, ZTF \citep{Masci18}, and ATLAS \citep{ATL14} to look for a possible similar periodicity or time scale. Actually all three surveys report this source, which is rather bright (V$\simeq$12.8). The time span covered is quite long, from MJD 57000 to now. Besides small offsets due to the different filters used by the surveys (V and g for ASAS-SN, g and r for ZTF, c and o for ATLAS) all the curves show a substantial agreement for the time intervals in common.

The star was first classified as a variable by ASAS-SN based on the first two years of observations in the Johnson\'s V band \citep{Jay18}.
In that catalog (VIZIER II/366) the star was formally identified with two sources with the same mean magnitude and amplitude, and with periods of 545 and 540 days, clearly being the same star.
Multiple formal identifications are also present in the Alerce (ZTF-based) (https://alerce.online/) alert database.
The ZTF may also be searched at the IRSA-ZTF website with a cone-search tool, giving coordinates and a radius (https://irsa.ipac.caltech.edu/cgi-bin/Gator/nph-dd). The ATLAS archive may be queried with a forced photometry tool (https://fallingstar-data.com/forcedphot/queue/) and can be explored using the SIMBAD identifier.

\section{Data analysis}

\begin{figure}[htbp]
\centering
\includegraphics[width=16cm]{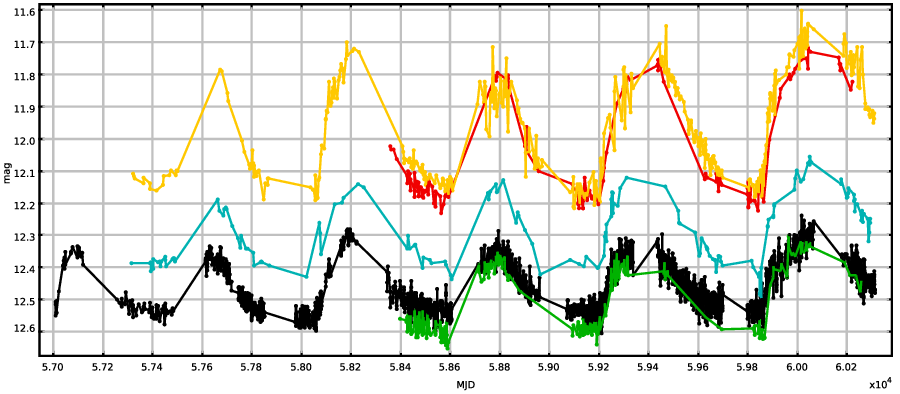} 
\caption{Light curve of IGR J06074+2205 for the all the datasets of Table 1.  Abscissa in MJD; ordinate in magnitudes. Orange=ATLAS-o; Cian=ATLAS-c; Red=ZTF-r; Green=ZTF-g; Black=ASASSN-V+g.}
\label{lcIGR}
\end{figure}

First of all we cleaned the downloaded light curves for data with large formal errors and made an average of multiple data points taken in the same night. Some obvious outliers were also excluded by visual inspection of the light curves. We recovered from the ASAS-SN archive the last 3000 days of observations, which allowed to have both V-band and g-band data of the source for 102 days. We could therefore derive a magnitude offset between V and g and build a single g-band light curve covering the time span since MJD 57000 to present.
  Fig. \ref{lcIGR} reports these light curves, after removing some bad quality data.

Then we performed a periodicity search using the FFT technique with the Period04 software \citep{Lenz05} for all the light curves, separately for each filter: the periods obtained are given in Table \ref{tab1}. One $\sigma$ uncertainties were computed with the Monte Carlo simulation provided by Period04, but are likely underestimates, as shown by the different periods found using the ZTF g and r bands.

\begin{table} 
\caption{Data sets used for period search.}
\vspace{3mm}  
\centering
\begin{tabular}{lccc}
\hline
	 Telescope & Time-span MJD & years & Period (d) \\
  \hline \hline
ASAS-SN-V & 57008 58451 & 2014-12-16 2018-11-28 & 549$\pm$3 \\
ASAS-SN-g & 58000 60300 & 2017-09-03 2023-12-21 & 655$\pm$2  \\
ZTF-g 	  &58000 60300  & 2017-09-03 2023-12-21 & 621$\pm$4  \\ 	  
ZTF-r 	  &58357 60221  & 2018-08-26 2023-10-03 & 633$\pm$4  \\ 	
ATLAS-c   &57314 60294  & 2015-10-18 2023-12-15 & 613$\pm$2  \\ 	
ATLAS-o	  &57319 60311  & 2015-10-23 2024-01-01 & 613$\pm$2  \\ 	
ASAS-SN-V+g &57000 60300 & 2014-12-16 2023-12-21 & 613$\pm$3  \\ 	
\hline
\end{tabular}\label{tab1}
\end{table}

\bigskip

\begin{figure}[htbp]
\centering
\includegraphics[width=16cm]{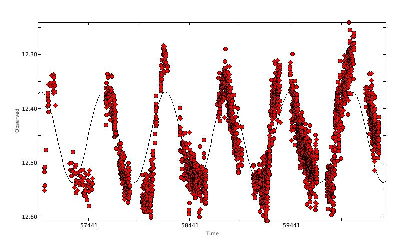} 
\caption{Light curve of IGR J06074+2205 for the ASAS-SN data including the early V-band converted to the g-band. The sinusoidal fit of 613 days is shown. Abscissa in MJD; ordinate in magnitudes.}
\label{asassn}
\end{figure}

\begin{figure}
\centering
\includegraphics[width=16cm]{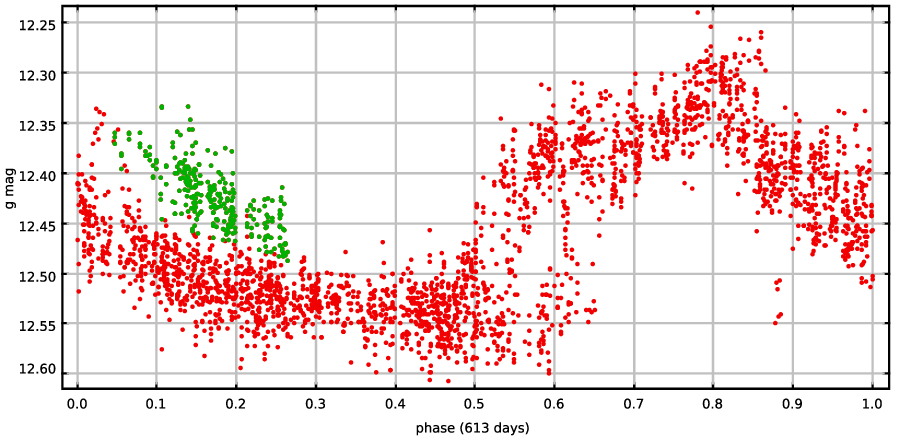} 
\caption{Phase-reduced light curve of IGR J06074+2205 for the ASAS-SN data shown in Fig.\ref{asassn} using a period of 613 days. The most recent data are marked in green.}
\label{faseIGR}
\end{figure}

As example, in Fig.\ref{asassn} we report the ASAS-SN V+g data, with superimposed the sinusoidal fit computed with Period04 of 613 days. In Fig.\ref{faseIGR} we report this light curve phase-reduced: the phase of each observation is computed taking into account the time and period, while the magnitudes are the observed ones, not the deviation from the sinusoidal fit of Fig.\ref{asassn}. In this plot it is apparent a group of data between phase 0.05 and 0.25, marked in green, which are above the bulk of the data. We have checked that this group is made just of data of the year 2023: this is apparent also from Fig.\ref{asassn} where the last descending branch of the light curve is above the sinusoidal line.
We remark that an FFT analysis of the first part of the light curve (sample ASAS-SN V) gives a period of 540 days, as reported in \citet{Jay18}: this short part of the light curve contains only the first 3 maxima, while the full set (ASAS-SN-V+g) contains 6 maxima.

A period of about 621 days was also reported by \citet{ATel16267} using the ZTF data only, in agreement with our ZTF-g sample.
 
It is apparent from Table \ref{tab1} and Figure \ref{lcIGR} that the star has a long term periodicity much longer than the X-ray outburst time scale of $\simeq$80 days. 
It is also evident from the ASAS-SN data in Fig.\ref{asassn} and \ref{faseIGR} that the light curve shape is not constant.

Considering Fig.\ref{lcIGR}, it is apparent that the variation amplitude is larger at longer wavelengths. This behaviour was already found by \citet{RF15} in their study of the long term optical luminosity of a sample of HMXB, for some stars of luminosity class III-V only. However they did not discuss the shapes of the light curves, but only the amplitudes in B,V,R,I Johnson-Cousins bands and their color trends. They also suggested that the source of the variable part of the optical luminosity must be the circumstellar disk, which is cooler than the Be stellar atmosphere and therefore produces a reddening of the total object luminosity when it becomes brighter.

\begin{figure}
\centering
\includegraphics[width=13cm]{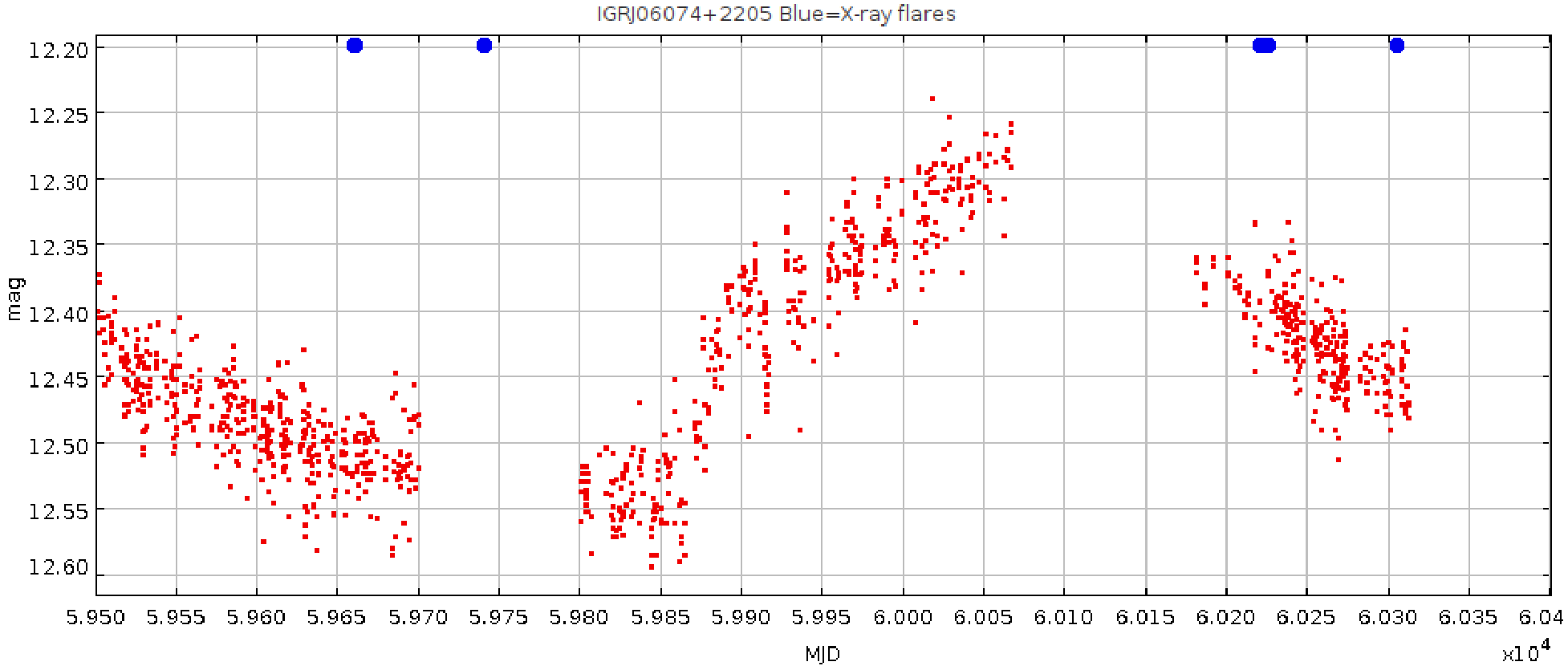} 
\caption{The recent light curve of IGR J06074+2205: blue dots mark the epochs of X-ray flares reported in various ATels.}
\label{xrayflares}
\end{figure}

Finally we report in Fig.\ref{xrayflares} the recent light curve of IGR J06074+2205 with indicated the dates of X-ray flares reported in ATels: maybe by chance, but all the flares happened when the optical luminosity was near the minimum or in the descending branch.

Given the discovery of a long term trend, we searched in the ASAS-SN, Alerce-ZTF and ATLAS archives all the HMXB with luminosity class III-V in the \citet{RF15} sample, to see if any other star showed a clear period or variability time scale in the light curve. 

 We report briefly for each of them a comment on the shape of the light curve.
 
4U 0115+63 (alias V635Cas): variable, with a strong episode of dimming, likely due to dust obscuration.

IGR J01363+6610: the ATLAS lightcurve may be fitted with a slow, small amplitude (0.05 mag) sinusoidal variation,  with a period $\simeq$1430 days, but with substantial scatter (0.04 mag rms).

IGR J01583+6713: slow decreasing trend, with small amplitude oscillations with period about 1100 days and rms scatter 0.05 mag;

RX J0240.4+6112: the ASASSN light curve shows a constant trend.

V0332+53 (alias BQ Cam): the source is a bit faint for ASASSN, the light curve shows a nearly constant trend with substantial scatter.  The ATLAS curve shows a clear decreasing and oscillating trend with a fast increase in the last year.

RX J0440.9+4431: the ASASSN curve shows a slow monotonic increasing trend without clear periodic oscillations. 

XTE J1946+274: the ATLAS light curve shows a decreasing trend in the first year, followed by a basically constant trend with significant scatter.

KS 1947+30: the ATLAS light curve shows a decreasing trend in the first year, then an oscillating behaviour with 0.1 mag amplitude and time scale of about 1 year, superposed to a long term trend.

EXO 2030+375: too faint for ASASSN; the ATLAS light curve show an increase in the first year, followed by a nearly constant behaviour, with long oscillations (period 1880 days) of small (0.03 mag) amplitude and 0.07 mag rms scatter. The period of 5 years is not much greater than the time interval covered by the observations, so it is just tentative.

GRO J2058+42: ATLAS shows a long term trend, increasing for some years, then decreasing, firstly fast and then slow, with some apparent short flares.

SAX J2103.5+4545: marked periodic oscillations with non sinusoidal shape due to long minima at nearly constant luminosity.

IGR J21343+4738: the ASASSN light curve shows a long term trend with substantial scatter: selecting data from a single camera, a small amplitude oscillating trend appears with time scale 735 days, but always with substantial scatter.

4U 2206+54: rather bright (mag 10); both the ASASSN and the ZTF database show puzzling data probably due to saturation problems. The ASASSN-V data however show a low amplitude oscillating behaviour with constant mean luminosity.

SAX J2239.3+6116: the ATLAS light curve shows a monotonic increasing trend without clear oscillations.

Overall all these stars show some level of variability: 6 stars show basically long term trends, 7 stars show some small amplitude oscillations, only 2 are rather constant. This is similar to what was found in the sample of HMXB in the Small Magellanic Cloud by \citet{Rajo11}.

Besides IGR J06074+2205, only one other star, SAX J2103.5+4545, shows variations of large amplitude, with a clear period of 420 days, and a "redder when brighter" behaviour.

\begin{figure}
\centering
\includegraphics[width=12cm]{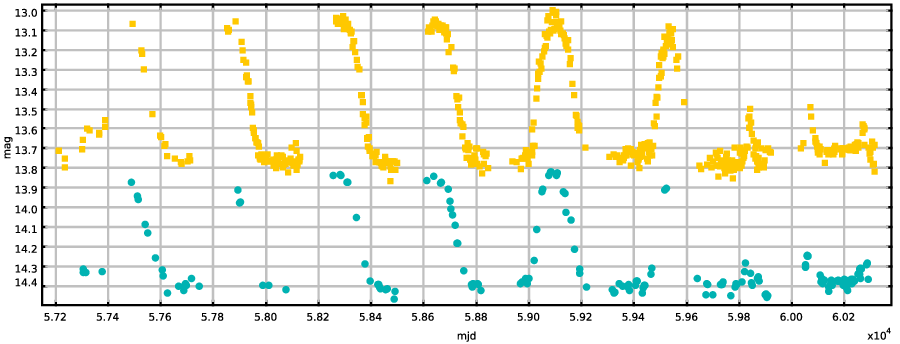} 
\caption{Light curve of SAX J2103.4+4545 with the ATLAS data. Orange = o band, cian = c band. The gap at 60000 might contain an optical peak.}
\label{sax2103}
\end{figure}

SAX J2103.5+4545 is a B0Ve X-ray binary with a neutron star and a high-mass stellar companion, at a distance of 6.2 kpc. It was discovered in 1997 by the BeppoSAX satellite and it has a spin period of 358.61$\pm$0.03 s \citep{Hu98}. It was reported on INTEGRAL/IBIS survey \citep{Bi16} as a variable High Mass X-ray Binary.
Its ATLAS light curve is shown in Fig.\ref{sax2103}. Both stars were among the five stars of the \citet{RF15} sample which showed some times the loss of their H-alpha emission line. The light curve of SAX J2103.5+4545 was studied also by \citet{AA} in the years 2015-2019 who found a period of 412 days not strictly constant, but did not discuss its color variation. Our study on a longer time interval confirms their period.

If the long term light curve of a HMXB is due to the circumstellar disk we would expect some correlation of the luminosity with the equivalent width (E.W.) of the H$\alpha$ emission. We recovered therefore from the paper by \citet{RF16} the E.W. measures of IGR J06074+2205, covering the time interval MJD 54031.5 to 56943.5. Unfortunately this time interval is not covered by the ASAS-SN monitoring, which starts from MJD 57008, nor by ATLAS or ZTF which start even later; however it is 2912 days long so that a periodicity of about 600 days should be detectable if present.

\begin{figure}
\centering
\includegraphics[width=12cm]{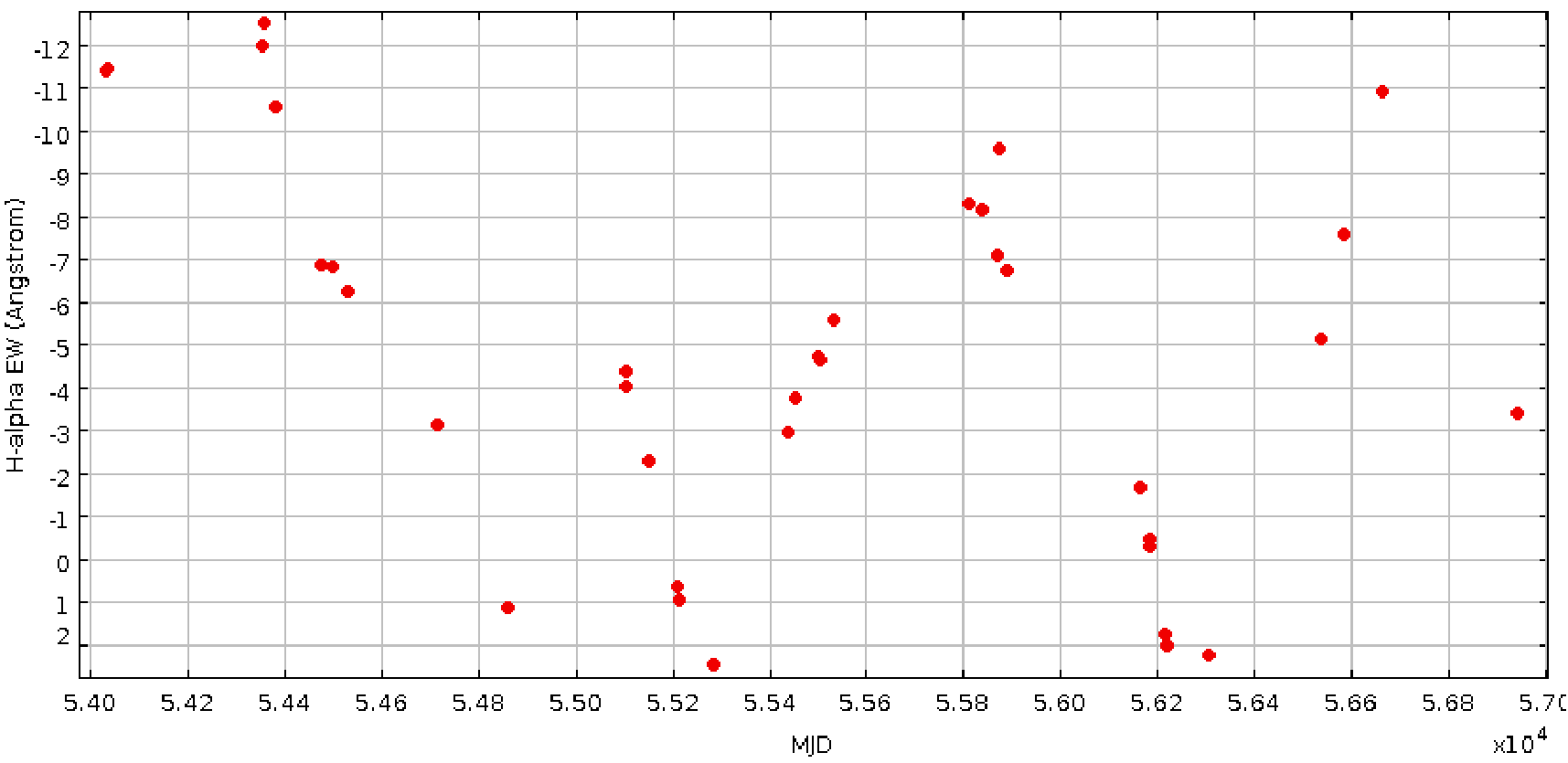} 
\caption{Light curve of the H$\alpha$ E.W. of IGR J06074+2205 with the data of \citet{RF16}.}
\label{ew}
\end{figure}

The E.W. light curve is shown in Fig. \ref{ew}, and shows a possible recurrent behaviour with four local maxima around MJD 54354, 55106, 55873, and 56665, the second one being much fainter than the others. If this interpretation were correct, then the H$\alpha$ maxima would occurr with a time scale of about 770 days, with their intensities depending on the amount of the mass loss. This time scale is however rather different from the period of the optical light curve, so we regard this interpretation as doubtful. 

We checked the present status of the H$\alpha$ line with the 50 cm telescope of the S. Lucia Observatory (MPC 589) with 8 Angstrom resolution from 2024-01-11 to 2024-03-12 , i.e. MJD 60382: the line was always in emission with E.W. $\simeq$ 11 Angstrom, its typical maximum value. If the time scale of 770 days quoted above were stable, the maximum should occurr around MJD 60515, in September 2024, while it seems that it has been already reached.

\begin{figure}
\centering
\includegraphics[width=13cm]{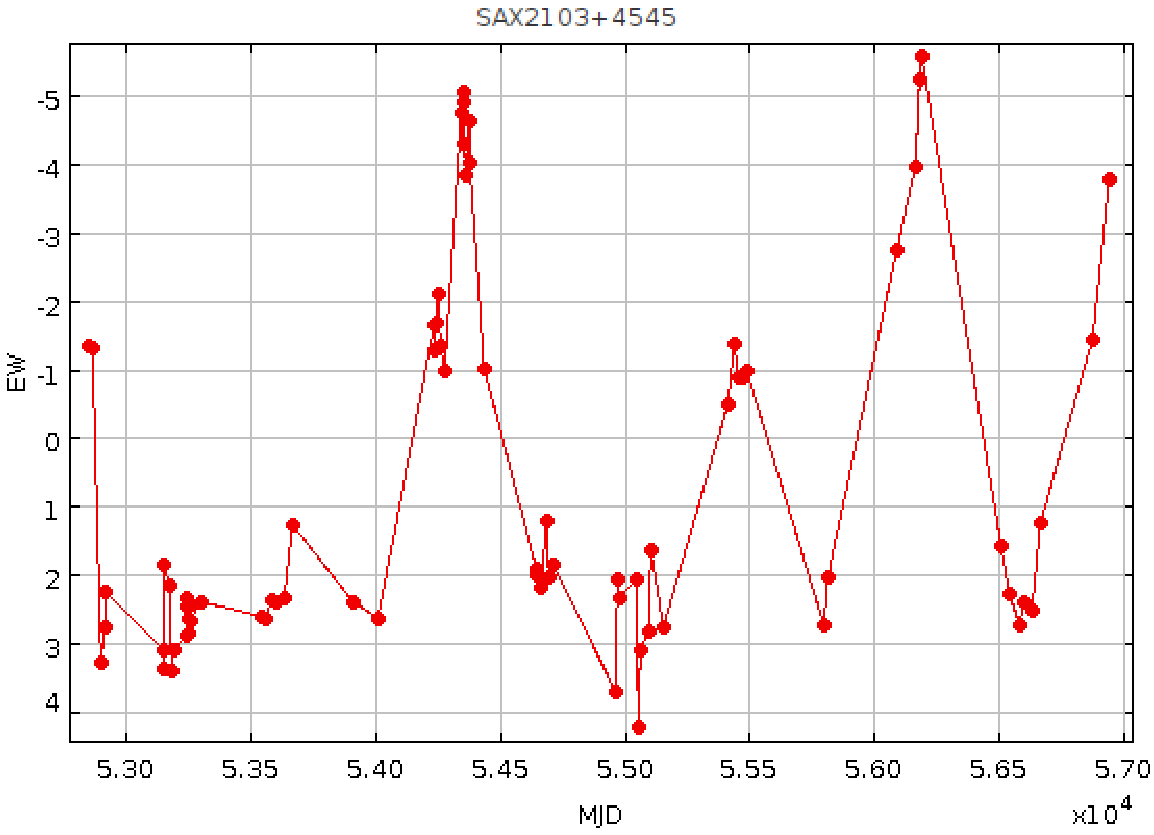} 
\caption{Time variation of the H$\alpha$ emission of SAX J2103.5+4545 from \citet{RF16}. E.W. in Angstrom.}
\label{saxew}
\end{figure}

 We then made a time analysis also of the H$\alpha$ E.W. of SAX J2103.5+4545, finding a somewhat recurrent behaviour, which is shown in Fig.\ref{saxew}. The line was often found in absorption (positive values) with some recurrent episodes of emission of different intensity, dominated by two maxima $\simeq$1800 days apart, so it cannot be well reproduced with a simple sinusoid. As for IGR J06074+2205, also this time interval is not a clear multiple of the optical period.

\section{Discussion}
A possible explanation of the optical variability of IGR J06047+2205 and SAX J2103.5 +4545 is the precession of the circumstellar disk of the Be star: the theorical reasons for the existence of a precession were recently discussed by \citet{MC24} and references therein.
The period of this precession is generally referred to in the literature as super-orbital period.
Such precession would be irrelevant for the optical luminosity if the disk were nearly face-on for the observer, because its surface would always be seen entirely; on the contrary, if the disk were nearly edge-on its apparent surface would be of variable extent and therefore could produce a variable apparent total flux. This could explain why only a small (2/16) fraction of HMXB shows large quasi-periodic optical variations. Given that the disk mass is somewhat variable due to the stellar wind irregularities, also the disk luminosity is not expected to be constant, and the light curve shape may be not very regular.
We remark that in this scenario the actual precession period of the disk would be the double of the observed light curve period.

Another explanation for the V-band superorbital variability was suggested by \citet{Alf17} from a detailed spectroscopic study of H 1145-619, namely the propagation of density waves in the Be star disk, as their characteristic time-scales are very similar to those of the V/R cyclic variability (see also \citet{Rajo11}.)

  A paper by \citet{TC20} shows a mild correlation in a log-log plane between orbital and super-orbital periods in a sample of HMXB not including our ones (their Fig.1). SAX J2103.5+4545, with a period of 12.5 days falls nicely on their relation either with a super-orbital period of 422 days or its double; also IGR J06074+2205 falls on the relation if the period of its light curve is half its precession period.

Given the very different time scales of the optical light curves and of the X-ray flares recurrence of IGR J06074+2205 (80 days) and SAX J2103.5+4545 (12.5 days, \citet{RF10}), it is apparent that the optical variability has little to do with the X-ray flares.

\section{Conclusions}\label{secconc}
Out of a sample of 16 High Mass X-ray Binaries with a Main Sequence Be star, only two show a marked quasi periodic optical light curve, with a large amplitude an evident "redder when brighter" behaviour.

The observed variations of the long term optical luminosity could be due to a geometrical effect, namely the precession of the circumstellar disk of the Be star with a period much longer than the orbital one. A different explanation is however possible, namely the propagation of a density wave in the disk of the Be star.

The X-ray flare dates of these two sources show no correlation with the optical luminosity, as expected if they are due to the crossing by the companion neutron star of the circumstellar disk.

\end{document}